\documentclass[10pt,conference]{IEEEtran}
\IEEEoverridecommandlockouts
\usepackage{cite}
\usepackage{amsmath,amssymb,amsfonts}
\usepackage{algorithmic}
\usepackage{graphicx}
\usepackage{textcomp}
\usepackage{xcolor}
\def\BibTeX{{\rm B\kern-.05em{\sc i\kern-.025em b}\kern-.08em
    T\kern-.1667em\lower.7ex\hbox{E}\kern-.125emX}}

\usepackage{tikz}
\usepackage{pgfplots}
\usepackage{listings}
\usepackage{pifont}
\usepackage{multirow}
\usepackage{xspace}
\usepackage{arydshln}
\usepackage[framemethod=tikz]{mdframed}
\usepackage{enumitem}
\usepackage{diagbox}
\usepackage{relsize}
\usepackage{colortbl}
\usepackage{rotating}
\usepackage{flushend}
\usepackage{hyperref}

\usetikzlibrary{arrows}

\usepackage{tcolorbox}
\definecolor{mycolor}{rgb}{0.122, 0.435, 0.698}
\definecolor{gray1}{gray}{0.3}

\definecolor{codegreen}{rgb}{0,0.6,0}
\definecolor{codegray}{rgb}{0.5,0.5,0.5}
\definecolor{codepurple}{rgb}{0.58,0,0.82}
\definecolor{backcolour}{rgb}{0.95,0.95,0.92}
\lstdefinestyle{mystyle}{
    commentstyle=\color{codegreen},
    keywordstyle=\color{magenta},
    numberstyle=\footnotesize\color{codegray},
    stringstyle=\color{codepurple},
    basicstyle=\footnotesize\ttfamily,
    breakatwhitespace=false,         
    breaklines=true,                 
    captionpos=b,                    
    keepspaces=true,                 
    numbers=left,                    
    numbersep=5pt,                  
    showspaces=false,                
    showstringspaces=false,
    showtabs=false,                  
    tabsize=2,
    columns=fixed
}
\newcommand{\result}[1]{%
\begin{tcolorbox}[colframe=mycolor,boxrule=0.5pt,arc=4pt,
      left=5pt,right=5pt,top=5pt,bottom=5pt,boxsep=0pt,width=\columnwidth]%
      {#1}      
\end{tcolorbox}%
}   

\definecolor{darkgreen}{rgb}{0.0, 0.5, 0.0}
\definecolor{darkred}{rgb}{0.82, 0.1, 0.26}

\newcommand{\todo}[1]{}%
\renewcommand{\todo}[1]{{\color{red} TODO: {#1}}}%

\newcommand{\me}{B\"{o}hme\xspace}

\newcommand{\rpm}{\sbox0{$1$}\sbox2{$\scriptstyle\pm$}
  \raise\dimexpr(\ht0-\ht2)/2\relax\box2 }

\newmdenv[innerlinewidth=0.5pt, roundcorner=4pt,linecolor=gray,innerleftmargin=0pt,
innerrightmargin=1pt,innertopmargin=2pt,innerbottommargin=0pt]{quotes}

\newif\ifblinded

\tikzset{
  treenode/.style = {align=center, inner sep=0pt, text centered,
    font=\sffamily},
  arn_n/.style = {treenode, circle, black, draw=black,
    text width=1.5em, minimum size=1.1cm},
  arn_r/.style = {treenode, circle, black, draw=gray1, 
    text width=1.5em, minimum size=1.1cm},
  arn_x/.style = {treenode, rectangle, gray1, draw=gray1,
    minimum width=0.9cm, minimum height=0.9cm}
}

\renewcommand{\IEEEbibitemsep}{0pt plus 0.5pt}
\makeatletter
\IEEEtriggercmd{\reset@font\normalfont\fontsize{7.9pt}{8.40pt}\selectfont}
\makeatother
\IEEEtriggeratref{1}

\begin{document}
\title{Statistical Guarantees for Automated Software Testing}
\title{Extrapolation in Software Testing}
\title{Guarantees in Software Testing}
\title{Assurance in Software Testing: A Roadmap}

\author{\IEEEauthorblockN{Marcel B\"{o}hme}
\IEEEauthorblockA{Monash University\\ \href{mailto:marcel.boehme@acm.org}{marcel.boehme@acm.org}\\[-0.3cm]}
}

\maketitle
\thispagestyle{plain}
\pagestyle{plain}

\begin{abstract}

As researchers, we already understand how to make testing more effective and efficient at finding bugs. However, as fuzzing (i.e., automated testing) becomes more widely adopted in practice, practitioners are asking: Which assurances does a fuzzing campaign provide that exposes no bugs? When is it safe to stop the fuzzer with a reasonable residual risk? How much longer should the fuzzer be run to achieve sufficient coverage?

It is time for us to move beyond the innovation of increasingly sophisticated testing techniques, to build a body of knowledge around the explication and quantification of the testing process, and to develop sound methodologies to estimate and extrapolate these quantities with measurable accuracy.
In our vision of the future practitioners leverage a rich statistical toolset to assess residual risk, to obtain statistical guarantees, and to analyze the cost-benefit trade-off for ongoing fuzzing campaigns. We propose a general framework as a first starting point to tackle this fundamental challenge and discuss a large number of concrete opportunities for future research.
\end{abstract}


\section{Introduction}
\vspace{0.3cm}
\result{\small\emph{Cognitive psychology tells us that the unaided human mind is vulnerable to many fallacies and illusions because of its reliance on its memory for vivid anecdotes rather than systematic statistics.}\\[0.1cm]%
\hspace*{\fill} --- Prof. Steven Pinker (Dep. of Psychology at Harvard)\vspace{-0.1cm}
}  
\subsection{A Vivid Anecdote}

Last year, we\,attended\,a\,meeting\,with several representatives of a large company that provides security assessment services for governments and industries worldwide. In preparation for this\,meeting, we\,learned\,about their\,product\,portfolio\,and found that their main product, a protocol-based blackbox fuzzer,\footnote{We shall use the terms \emph{fuzzing} and \emph{automated testing} interchangeably.} can be used for \emph{security certification of medical devices} (IEC 62443-4-2).\footnote{The assessment scheme for IEC 62443, the worldwide standard for security of Industrial Control Systems, is operated by the ISA Security Compliance Institute and offered within the Embedded Device Security Assurance (EDSA) product which ascertains compliance with IEC 62443-4-2 \cite{iec}.} While the fuzzer is part of a larger certification process,\footnote{The ISASecure EDSA certification also requires that the organization follows a robust, secure software development process and that the product has properly implemented the security-related functional requirements.} it is primarily the fuzzer's task to identify vulnerabilities that could be exploited remotely over the network. 

Now, medical devices are safety-critical systems and undetected vulnerabilities can be life threatening. For instance, Halperin et al. \cite{medDev} describe several attacks on an implantable cardioverter defibrillator to control when electrical shocks are administered to the patient's heart. Hence, subjecting medical devices to rigorous cyber security assessment is a powerful mitigator of cyber risks.
This inspires \emph{trust} in the certificate. 
A violation of this trust (e.g., if an attacker exploited an undetected vulnerability in a medical device that is \emph{certified}) would be disastrous. Thus, the company's reputation and the certificate's credibility depend, at least in part, on the \emph{assurances} which the fuzzer provides.

We walked into that meeting wondering about this question. Which assurances are derived for the certificate from applying the company's fuzzer?
To paraphrase Djikstra, fuzzing can be used only to show the presence of vulnerabilities, not their absence.
How do they effectively assess the residual risk of a fuzzing campaign that finds no vulnerabilities? How do they arrive at the decision to stop the fuzzer and to proceed with the certification?
It turns out that the certification scheme does not specify how much fuzzing is sufficient for certification. Neither does it specify a concrete value for the \emph{allowable} residual risk that an undiscovered vulnerability still exists. In practice, \emph{the decision is with the individual security researcher}. Even if a concrete threshold value was specified, there is no statistical framework available that would allow the researcher to quantify the residual risk for an ongoing campaign \cite{harrold,bertolino,whalen}.

In the end, we were told, the decision is mostly based on \emph{experience}. Clearly, a long-running fuzzing campaign provides much stronger assurances than a shorter one; meaning, the residual risk decreases as the length of the campaign increases. Hence, intuitively there is a particular point in time when it is both economical and safe to abort a fuzzing campaign that has found no vulnerabilities. 

\subsection{Call for Systematic Statistics}
In this paper, we argue that we ought to do better than relying on an individual's experience. The security researcher should be able to systematically assess and quantify the inherent uncertainty. The certification authority should be able to provide concrete guidance in the form of measurable threshold values. 
This offers an opportunity for us as software engineering \emph{researchers} to develop a rich statistical toolset that will enable software engineering \emph{practitioners} to assess and quantify the automated testing process. 

Senior members of our community have previously called for a general statistical framework. 
In 2000, Harrold \cite{harrold} established the ``development of techniques and tools for use in \emph{estimating, predicting, and performing testing}'' as a key research objective for future software testing research. Seven years later,  Bertolino \cite{bertolino} corroborated that ``we will need to make the process of testing more effective, \emph{predictable} and effortless''. Yet today, Whalen \cite{whalen} observes ``there is \emph{no sound basis to extra\-polate} from tested to untested cases''.\footnote{In the quotes, the emphasis in \emph{italic letters} is mine.}

Over the last couple of years, fuzzing has become widely adopted in practice, including by Google \cite{oss}, Microsoft \cite{springfield}, Adobe \cite{adobe}, Mozilla \cite{mozillaFuzzing}, and Facebook \cite{open1}. 
Argueably, we already know how to \emph{control, bias, and optimize} the fuzzing process so as to make it more efficient \cite{efficiency,efficiency2}. Search-Based Software Testing (SBST) has become a mature field of research \cite{sbst}. Practitioners utilize a large collection of extremely efficient fuzzing strategies.

Going forward, there needs to be a similarly rich body of work that facilitates a deeper understanding how to \emph{quantify, measure, and assess} the fuzzing process so as to make it more explicable. Only then, practitioners will be able to answer very practical questions, such as

\vspace{-0.2cm}
{\small%
\begin{enumerate}[leftmargin=0.5cm, itemsep=4pt]
  \item \textbf{Residual risk}. Supposing no vulnerability has been found after generating $n$ test inputs, what is the probability to generate a new test input that does expose a vulnerability? More generally, \emph{how much has been learned about the program's behavior, and how much more remains to be learned?} 
  \item \textbf{Cost-benefit Trade-off}. Supposing no vulnerability has been found after generating $n$ test inputs, how  does the residual risk decrease if time is spent generating $m$ more test inputs? More generally, \emph{how much more could be learned about the program's behavior within an \emph{additional} time budget?}
  \item \textbf{Testability}. How ``difficult'' is it for an arbitrary fuzzer to discover vulnerabilities in the program? More generally, \emph{what is the average amount of information that the program reveals about its behaviors per generated test input?}
  \item \textbf{Effectiveness}. How effective is the fuzzer w.r.t. discovering  vulnerabilities? More generally, \emph{what is the \emph{asymptotic} number of (discrete) program behaviors that the fuzzer is able to expose?}
\end{enumerate}}
\vspace{-0.2cm}

So then, why have previous calls of eminent researchers for extrapolation in software testing gone unheeded? To most of us, it seems counter-intuitive that one can soundly extrapolate from tested to untested program behaviors. Rushby (1992) \cite{rushby2} explains by analogy: A physical system exhibits behavior that is \emph{continuous}, which allows engineers to test a few critical points of a bridge to make predictions about the safety of the whole bridge. In contrast, a software system exhibits behavior that is \emph{discretized} and \emph{discontinuous}. Hence, ``tests provide information on only the state sequences actually examined; without continuity there is little reason to suppose the behavior of untested sequences will be  close  to tested ones,  and therefore little justification for extrapolating from tested cases to untested ones`` \cite{rushby2}. However, in the last 16 years statistics has made huge strides, providing statistical tools that make \emph{no assumptions about continuity} (here, of program behaviors).

Moreover, many of us with a formal methods background are swayed by counter-examples. It is trivial to construct a small program with an \texttt{if}-condition where the \texttt{else}-branch is evaluated with an infinitesimal probability. How can one ever expect to extrapolate in a sound manner? The answer is two-pronged: (i)~It is possible to make \emph{no assumptions} about the probability of some behavior to be observed or how many behaviors there are. (ii)~Empirically, the typical program does not resemble such counter-examples. Indeed, an empirical evaluation of the performance of several biostatistical estimators for software testing showed tremendous promise \cite{stads}.

\section{A General Statistical Framework for Software Testing: Challenges and Oportunities}
Estimation and extrapolation are classical problems in statistics. We can understand testing essentially as a sampling of program behaviors. As more tests are executed, we learn more about the program's behaviors. Our confidence in the program's correctness increases. To quantify this confidence (and generally the uncertainty about any concrete statements made, given only limited data) is a fundamental building block of statistics. Hence, it stands to reason to construct a \emph{general statistical framework} for software testing.

An interesting initial attempt in this direction is the STADS framework \cite{stads}. \me borrows several innovative biostatistical methodologies from ecology to answer those practitioners' questions (on the left) about residual risk, cost-benefit, testability, and effectiveness. The ecologic analogy is very powerful. It addresses Rushby's concerns \cite{rushby2} about discontinuity and provides direct access to over thirty years of research in ecological biostatistics. However, while STADS is an excellent starting point, the author introduces only a limited number of estimators and leaves many questions unanswered. Major hurdles are still ahead of us, providing abundant opportunities for future research.

It is also the purpose of this article to highlight these peculiarities, to discuss concrete  opportunities for future work, and to elucidate the underlying assumptions, their impact on the statistical guarantees, and how they can be addressed. These are the construction sites  on our path to a fundamental understanding of the testing process.

\subsection{Software Testing as Species Discovery}
%
%


Ecology is concerned with species discovery. For instance, to study the species diversity of arthropods in a tropical rain forest, ecologists would first sample a large number of individuals from that forest and determine their species. The species in the sample are said to be \emph{discovered}. Exhaustive sampling is often prohibitive\footnote{For instance, it took 102 ecology researchers 66 person-years to sample 129,494 arthropod individuals representing 6144 species from 0.48 ha of tropical rain forest \cite{tropical}.} and the discovered species represent only a proportion of all species. Hence, the fundamental challenge is to \emph{extrapolate from the species observed in the sample}. Biostatisticians spent the last three decades \cite{incidenceSurvey} developing a framework that addresses this extrapolation challenge in ecology. 

\me's \cite{stads} key observation was that testing is about \emph{discovery}, as well. For instance, let us call each branch that is covered by an input, a \emph{species of that input}. To maximize branch coverage, a fuzzer would first sample a large number of  test inputs from that program's input space and determine their species. The species in the sample are said to be discovered. Again, exhaustive sampling is prohibitive and the discovered species (here, the covered branches) represent only a proportion of all species (here, all feasible branches). Again, the fundamental challenge is to \emph{extrapolate from the species observed in the sample of generated test inputs}.

The species for an input can be defined in several ways, depending on the dynamic program properties of interest. We could say a new species is discovered when the fuzzer generates an input that
\begin{itemize}
  \item exercises a statement, branch, path, or any other program element not previously exercised (code coverage),
  \item kills a mutant not previously killed (mutation-adequacy),
  \item exposes a previously unexposed assertion violation, program crash, memory error, non-functional error, race condition, etc. (bug finding),
  \item discovers a previously undiscovered sensitive information flow \cite{mutaflow},
  \item or any other discrete property of the program's behavior,
  \item or any combination of the above.
\end{itemize}
\vspace{0.2cm}

Within the ecologic analogy, a \emph{test input} is the individual or sampling unit, and the observed program behavior is the input's species. An input can belong to multiple \emph{species}. For instance, each statement that an input executes may be considered a species of that input. A \emph{fuzzer} samples test inputs and discovers species. A \emph{dynamic analysis} identifies an input's species. For instance, the \texttt{gcc} tool identifies the statements exercised by an input.

The perspective of Software Testing and Analysis as Discovery of Species (STADS) provides direct access to 30+ years of research in ecological biostatistics \cite{incidenceSurvey}. In the following, we present examples of biostatistical estimators and their utility in automated testing.\footnote{Due to the lack of space, the reader is referred to \me \cite{stads} for a formal introduction of the STADS framework and the statistical models underpinning the framework.}

\textbf{Residual risk} can be assessed using the probability to discover a new species.\footnote{In ecology, discovery probability is called sample coverage and gives the proportion of individuals in the population belonging to a species represented in the sample.} The \emph{discovery probability} is the probability that the next generated test input leads to the discovery of a previously unseen species. 
The discovery probability $U(n)$ can be estimated accurately and efficiently for arbitrary species abundance distributions \cite{consistent,goodNormal,goodError, goodRobust} using Good-Turing \cite{good}
\begin{align}
\hat U(n) = \frac{f_1}{n}\\[-0.25cm]\nonumber
\end{align}
where $f_1$ is the number of (singleton) species having been observed exactly once in the campaign.\footnote{Intuitively, even after $n$ test inputs have been generated there are still $f_1$ species that have been observed exactly once; clearly, we can expect it takes at least as many test inputs to observe one of $f_0$ undiscovered species that have not been observed, at all.} Now, a test input can belong to some species and \emph{not} expose an error, or belong to the same species and expose an error. For the purpose of discussion let the latter be a species on its own. If no error has been exposed throughout the campaign, the Good-Turing estimator gives an upper bound on the probability to generate a test input that exposes an error.

\textbf{Cost-benefit trade-offs} can be assessed using statistical extrapolation. Given $S(n)$ species have been discovered after $n$ test inputs have been generated, we can estimate the number of species $S(n + m^*)$---that we can expect to discover if $m^*$ more test inputs are generated---using the following estimator \cite{shen,incidenceSurvey}
\begin{align}
\hat S(n+m^*) = S(n) + \hat f_0\left[1-\left(1-\frac{f_1}{n\hat f_0 + f_1}\right)^{m^*} \right]\\[-0.25cm]\nonumber
\end{align}
where $f_1$ is the number of (singleton) species having been observed exactly once in the campaign, and $\hat f_0=\hat S - S(n)$ is an estimate of the number of (undiscovered) species. If unknown, the asymptotic total number of species $S$ can be estimated as follows \cite{chao1,chao2}\vspace{0.3cm}
\begin{align}
\hat S = \begin{cases}
S(n) + f_1^2/(2f_2) & \text{ if } f_2>0\\
S(n) + f_1(f_1-1)/2 & \text{otherwise}
\end{cases}\label{eqn:s}\\[-0.25cm]\nonumber
\end{align}
where $f_2$ is the number of (doubleton) species that have been observed exactly twice throughout the campaign. Given a discovery probability $U(n)$ after $n$ test inputs have been generated, we can extrapolate the residual risk---in the case that $m^*$ more test inputs were generated---as follows \cite{sampleSurvey,incidenceSurvey}
\begin{align}
\hat U(n + m^*) = \frac{f_1}{n}\left(\frac{n\hat f_0}{n\hat f_0 + f_1}\right)^{m^*+1}\\[-0.25cm]\nonumber
\end{align}

\textbf{Effectiveness}. If the total number of species $S$ is unknown, it can be estimated as given in Equation~(\ref{eqn:s}). For instance, if one species corresponds to one mutant in mutation testing, then the $\hat S$ estimates the asymptotic number of mutants that \emph{can} be killed. Some mutants are stubborn while others cannot be killed at all \cite{stubborn}. $\hat G(n) = S(n)/\hat S$ gives an estimate of the \emph{feasible} mutation coverage.

\textbf{Generality}. The STADS framework \cite{stads} is \emph{general} in the sense that it does not depend on a specific programming language, or execution environment, testing objective, or test generation technique. For instance, STADS facilitates the extrapolation of statement coverage, mutation adequacy, or the number of bugs exposed for Java, C, and Python programs on Linux, Windows, and MacOS using CSmith \cite{csmith}, Randoop \cite{randoop}, or AFL \cite{afl,aflfast,aflgo,aflsmart}. 
%
%

\subsection{Flaky Tests and Non-Deterministic or Stateful Programs}
A test is \emph{flaky} if executing it twice on the same program may yield different outcomes. A flaky test may pass nine out of ten times but then fail for no obvious reason. Such a failure may or may not be spurious. Recently, Harman and O'Hearn \cite{open1}, reflecting on their experience at Facebook, established the presence of flaky tests as a key challenge in software testing research. In fact, researchers should assume that All Tests Are Flaky (ATAF). Statistics is well-equipped to deal with Harman and O'Hearn's ATAF-istic world. 

We say that a test is flaky due to \emph{non-determinism} if the probability of a certain outcome does not change substantially during testing. More specifically, the random outcomes of a flaky test should originate from the same probability distributions, and be mutually independent. In statistics, this is called an independent and identically distributed (\emph{IID}) random variable. Examples are concurrent programs where a particular thread schedule determines the test outcome, or probabilistic programs where the specified branching probabilities determine the test outcome. 

We say that a test is flaky due to \emph{statefulness} if the outcomes are not \emph{IID}. The outcomes of a flaky test are either mutually dependent (i.e., induce state changes) or come from different distributions (e.g., due to externally induced state changes). Regardless, there is some underlying state in the program or in its environment that changes in-between test executions. Examples are interactive programs with user-interfaces, such as Android apps or web apps. Executing the same event sequence twice may produce different outputs. For instance,  entering a sequence on a calculator program once ('1+1') or twice ('1+11+1') will give different results. 

Statistics provides a large set of methodologies for estimation and extrapolation in the presence of tests that are flaky due to non-determinism. In applied statistics, it is a common (though often an unrealistic) assumption that a random variable is \emph{IID} as it drastically simplifies the underlying statistical theory. To determine whether a flaky test is due to non-determinism (i.e., whether its outcomes are~\emph{IID}), we suggest the \emph{turning point test}. Existing statistical methodologies should be identified and studied empirically. For instance, to study the performance of an extrapolation methodology in the presence of flaky tests, we can simply compare the predicted value to the actual (future) value in several experiment repetitions.

In order to ``unflake'' tests that are flaky due to statefulness, we suggest to execute each test case on the \emph{exact same} state. The advent of advanced virtualization and containerization technology allows to capture, control, and restore the entire state of the program and all of its environment, including the entire operating system and the (virtual) machine it is running upon. Clearly, this removes the need to handle tests that are flaky due to statefulness within our statistical framework. 
%
%
%
%

\subsection{Extremely Rare and Rather Extreme Program Behaviors}
In software testing, we are often most interested in program behaviors that are both extreme and rarely observable, such as a program crash or a missed deadline in a real-time system. In applied statistics, \emph{rare event analysis} \cite{rare} and \emph{extreme value theory} \cite{evt} have become substantial fields of research with important applications, e.g., in economics, meteorology, actuarial science, physics, and ecology. For instance, Glasserman et al. \cite{rare1} propose an adaptive sampling method for the efficient estimation of the probability of a rare event occuring. Software engineering researches should study such highly innovative sampling methods, tailored for the discovery of rare events, to develop more efficient software testing techniques, or to estimate and extrapolate the residual risk that a vulnerability exists if none has been found.

In software testing, interesting behaviors can be clustered in several very narrow regions of the program's input space. For instance, some file formats start with a ``magic number'' \cite{thinair}; only if that magic number is correct will interesting program behaviors be exercised. In ecology, a large number of \emph{endemic species} may be clustered on remote islands. To facilitate estimation in the presence of endemic species, ecological biostatisticians leverage \emph{sample coverage-based estimators} \cite{ace,ice} or more generally \emph{Good-Turing  theory} \cite{goodtheory}. In future, existing biostatistical estimators should be evaluated empirically and specifically tailored to the context of software testing. There are particularly well-suited adaptive cluster sampling methods for efficient estimation in the presence of a large number of very rare and endemic species \cite{conroy,thomson}.

%

\subsection{Uncertainty about Correctness and the Oracle Problem}
The correctness of any statistical guarantee depends on the \emph{soundness of the dynamic program analysis} which distinguishes correct from incorrect program  behavior for a given input.
More specifically, we cannot assume that \emph{all} inputs that expose a vulnerability are recognized as such \cite{oracle}. Without a vulnerability-specific dynamic analysis, the fuzzer is unable to automatically \emph{recognize}, e.g., a privilige escalation attack, even if it is actually triggered by a crafted test input (i.e., an exploit). In practice, the statistical guarantees hold only with respect to the kinds of vulnerabilities that \emph{can} be recognized (e.g., a buffer overflow by ASAN \cite{asan}). It is interesting to note that a similar challenge exists in software verification where a program can be proven correct only modulo the specification.

Elbaum and Rosenblum \cite{uncert} discuss the problem of \emph{uncertainty} in software testing. When Amazon recommends to buy Sartre's ``Being and Nothingness'', Spotify recommends to listen to Bach's ``Goldberg variations'', or the GPS is slightly off by a few meteres, we can establish their correctness only to some degree of accuracy and with some certainty. Statistics is well-suited for assessing such uncertainties and inaccuracies. This allows to develop techniques that account for the possibility that the oracle (rather than the program) could be incorrect (i.e., cannot be defined with absolute accuracy).

\subsection{Vulnerabilities Outside the Fuzzer's Search Space}
Not all detectable vulnerabilities in the program may be within the fuzzer's search space. More specifically, even if all inputs that expose a vulnerability are recognized as such, we cannot assume that the fuzzer can generate them. For instance, suppose a vulnerability in an Android app is exposed only via system-level events (e.g., a low battery-level); if the fuzzer cannot generate system-level events, this vulnerability is clearly out of scope for the fuzzer. As such, an estimate of the residual risk quantifies how likely it is that \emph{this} Android fuzzer (or an attacker with a similarly powerful Android fuzzer) generates an event sequence that exposes an undiscovered (but discoverable) vulnerability.

In practice, any statistical guarantees are correct only w.r.t. the detectable vulnerabilities that are within the fuzzer's search space. In order to provide more general statistical guarantees, we should investigate (i) a sound extrapolation from the fuzzer's search space to the entire input domain \cite{coleman} and (ii) a sound integration of several individual estimates---from fuzzers which cover disjoint domains in the program's input space---into a single, global estimate.

Moreover, we argue that a fuzzer's effectiveness is defined by the \emph{asymptotic} number of vulnerabilities the fuzzer is able discover. We should leverage statistical methodologies \cite{chao1} to extrapolate from the number of vulnerabilities seen throughout a (non-exhaustive) fuzzing campaign to the asympotic total number of vulnerabilities, in order to derive a \emph{sound estimate of fuzzer effectiveness}. For a sound comparison of two different fuzzers, we should study how to efficiently quantify the overlap of their search spaces.

%
%

\subsection{Challenges of Adaptive Bias in Automated Test Generation}
Most search-based software testing (SBST) techniques introduce an adaptive bias that needs to be corrected in the statistical analysis. As discussed earlier, the statistical analysis is substantially simplified if we can assume that the outcome of every test the fuzzer generates is \emph{IID}. This is assumption indeed holds for all blackbox fuzzers that do not adapt based on feedback from the program. However, this assumption does not hold for SBST techniques, where the probability to discover a vulnerability is supposed to increase with the length of the fuzzing campaign. The fitness of previously generated test inputs will adaptively drive the fitness (or quality) of test inputs generated later in the fuzzing campaign. 

%
In ecology, there exist several strategies to correct such adaptive bias a postiori. Like in software testing, ecologists are interested in boosting the efficiency of species discovery. For instance, the adaptive bias in \emph{adaptive cluster sampling} \cite{cluster} is corrected using the \emph{Horvitz-Thompson estimator} \cite{horvitz}. For search-based testing, similar bias correction strategies should be developed and empirically evaluated. Unlike in ecology, we can measure the adaptive bias \emph{during} the campaign itself---by comparing predictions to the actual values. This may allow us to dynamically adjust for the bias.

An empirical investigation of the correlation of estimates across various fuzzers for same-length  campaigns would bring insights about the estimates' generality.
The program's source code and program binary provide an additional source of information that can be used to improve estimator performance. In future, the dependence of estimator bias and precision on the discovery probability \cite{good} can be investigated to develop better \emph{bias-correction} mechanisms.
%


\vspace{-0.05cm} 
\section{Conclusion}
Coming back to our earlier example: We \emph{can} do better than relying on an individual's experience when assessing the assurances of automated testing. The \emph{security assessor} can use Good-Turing \cite{good} to estimate (as residual risk) the probability that an attacker---with similar or less resources and a fuzzer with similar or less efficiency---discovers a vulnerability that the assessor did not discover. The assessor can leverage extrapolation methodologies \cite{incidenceSurvey} to predict the reduction of residual risk if there was time to generate $m$ more test inputs.
The \emph{certification company} can provide concrete guidance by setting different threshold values for different levels of allowed residual risk. The \emph{security certificate} can provide statistical guarantees and clearly state the conditions under which they hold. 

Yet, there are many challenges still ahead of us. \emph{Generality} of the assurances should be improved by developing statistical methodologies for cases where the current assumptions do not hold. \emph{Estimator performance} should be improved by accounting for the peculiarities of software testing. \emph{Empirical studies} of the performance of various estimators should be conducted for different programs, fuzzers, and oracles. \mbox{This is our vision.}


\begin{thebibliography}{10}
\providecommand{\url}[1]{#1}
\csname url@samestyle\endcsname
\providecommand{\newblock}{\relax}
\providecommand{\bibinfo}[2]{#2}
\providecommand{\BIBentrySTDinterwordspacing}{\spaceskip=0pt\relax}
\providecommand{\BIBentryALTinterwordstretchfactor}{4}
\providecommand{\BIBentryALTinterwordspacing}{\spaceskip=\fontdimen2\font plus
\BIBentryALTinterwordstretchfactor\fontdimen3\font minus
  \fontdimen4\font\relax}
\providecommand{\BIBforeignlanguage}[2]{{%
\expandafter\ifx\csname l@#1\endcsname\relax
\typeout{** WARNING: IEEEtran.bst: No hyphenation pattern has been}%
\typeout{** loaded for the language `#1'. Using the pattern for}%
\typeout{** the default language instead.}%
\else
\language=\csname l@#1\endcsname
\fi
#2}}
\providecommand{\BIBdecl}{\relax}
\BIBdecl

\bibitem{iec}
``{Security for industrial automation and control systems - Certification of
  IACS supplier security policies and practices},'' International
  Electrotechnical Commission, International Standard, 2015.

\bibitem{medDev}
D.~Halperin~et al., ``Pacemakers and implantable cardiac defibrillators:
  Software radio attacks and zero-power defenses,'' in \emph{IEEE Symposium on
  Security and Privacy}, ser. S\&P '08, 2008, pp. 129--142.

\bibitem{harrold}
M.~J. Harrold, ``Testing: A roadmap,'' in \emph{Proceedings of the Conference
  on The Future of Software Engineering}, ser. ICSE '00, 2000, pp. 61--72.

\bibitem{bertolino}
A.~Bertolino, ``Software testing research: Achievements, challenges, dreams,''
  in \emph{Future of Software Engineering}, ser. FOSE'07, 2007, pp. 85--103.

\bibitem{whalen}
Website, ``Lockheed martin webinar series: Michael whalen on the future of
  verification and validation.''
  \url{https://www.computer.org/cms/Computer.org/webinars/lmco/012413Slides-Whalen.pdf},
  Jan. 2013, accessed: 2017-05-13.

\bibitem{oss}
------, ``{OSS-Fuzz: Five Months Later},''
  \url{https://testing.googleblog.com/2017/05/oss-fuzz-five-months-later-and.html},
  2017, accessed: 2017-11-13.

\bibitem{springfield}
------, ``Microsoft: Project springfield,''
  \url{https://www.microsoft.com/Springfield/}, 2017, accessed: 2017-11-13.

\bibitem{adobe}
------, ``Security @ adobe.''
  \url{https://blogs.adobe.com/security/tag/fuzzing}, 2017, accessed:
  2017-11-13.

\bibitem{mozillaFuzzing}
------, ``Mozilla: Fuzzing firefox with peach.''
  \url{https://wiki.mozilla.org/Security/Fuzzing/Peach}, 2017, accessed:
  2017-11-13.

\bibitem{open1}
M.~Harman and P.~O'Hearn, ``From start-ups to scale-ups: Opportunities and open
  problems for static and dynamic program analysis,'' 2018, keynote at the 18th
  IEEE International Working Conference on Source Code Analysis.

\bibitem{efficiency}
M.~B\"{o}hme and S.~Paul, ``A probabilistic analysis of the efficiency of
  automated software testing,'' \emph{IEEE Transactions on Software
  Engineering}, vol.~42, no.~4, pp. 345--360, April 2016.

\bibitem{efficiency2}
------, ``On the efficiency of automated testing,'' in \emph{Proceedings of the
  22nd ACM SIGSOFT International Symposium on the Foundations of Software
  Engineering}, ser. FSE 2014, 2014, pp. 632--642.

\bibitem{sbst}
P.~McMinn, ``Search-based software test data generation: A survey: Research
  articles,'' \emph{Journal of Software Testing, Verification and Reliability},
  vol.~14, no.~2, pp. 105--156, Jun. 2004.

\bibitem{rushby2}
J.~Rushby, ``Formal methods and digital systems validation for airborne
  systems,'' NASA contractor report ; NASA CR-4551., Tech. Rep., 1993.

\bibitem{stads}
M.~B\"{o}hme, ``{STADS}: Software testing as species discovery,'' \emph{ACM
  Transactions on Software Engineering and Methodology}, vol.~27, no.~2, pp.
  7:1--7:52, Jun. 2018.

\bibitem{tropical}
Y.~Basset, L.~Cizek, P.~Cuenoud, R.~Didham, F.~Guilhaumon, O.~Missa,
  V.~Novotny, F.~\O{}degaard, T.~Roslin, J.~Schmidl, A.~K.~Tishechkin,
  N.~N.~Winchester, D.~Roubik, H.-P. Aberlenc, J.~Bail, H.~Barrios, J.~Bridle,
  G.~Casta\~{n}o, B.~Corbara, and M.~Leponce, ``Arthropod diversity in a
  tropical forest,'' \emph{Science}, vol. 338, no. 6113, pp. 1481--1484, 2012.

\bibitem{incidenceSurvey}
A.~Chao and R.~K. Colwell, ``Thirty years of progeny from chao's inequality:
  Estimating and comparing richness with incidence data and incomplete
  sampling,'' \emph{Statistics and Operations Research Transactions}, vol.~41,
  no.~1, pp. 3--54, 2017.

\bibitem{mutaflow}
B.~Mathis, V.~Avdiienko, E.~O. Soremekun, M.~B\"{o}hme, and A.~Zeller,
  ``Detecting information flow by mutating input data,'' in \emph{Proceedings
  of the 32nd IEEE/ACM International Conference on Automated Software
  Engineering}, ser. ASE, 2017, pp. 263--273.

\bibitem{consistent}
A.~B. Wagner, P.~Viswanath, and S.~R. Kulkarni, ``Strong consistency of the
  good-turing estimator,'' in \emph{Proceedings of the 2006 IEEE International
  Symposium on Information Theory}, 2006, pp. 2526--2530.

\bibitem{goodNormal}
C.-H. Zhang and Z.~Zhang, ``Asymptotic normality of a nonparametric estimator
  of sample coverage,'' \emph{The Annals of Statistics}, vol.~37, no.~5A, pp.
  2582--2595, 10 2009.

\bibitem{goodError}
H.~E. Robbins, ``Estimating the total probability of the unobserved outcomes of
  an experiment,'' \emph{The Annals of Mathematical Statistics}, vol.~39,
  no.~1, pp. 256--257, 02 1968.

\bibitem{goodRobust}
A.~Orlitsky and A.~T. Suresh, ``Competitive distribution estimation: Why is
  good-turing good,'' in \emph{Proceedings of the 28th International Conference
  on Neural Information Processing Systems}, ser. NIPS'15, 2015, pp.
  2143--2151.

\bibitem{good}
I.~J. Good, ``The population frequencies of species and the estimation of
  population parameters,'' \emph{Biometrika}, vol.~40, pp. 237--264, 1953.

\bibitem{shen}
T.-J. Shen, A.~Chao, and C.-F. Lin, ``Predicting the number of new species in
  further taxonomic sampling,'' \emph{Ecology}, vol.~84, no.~3, pp. 798--804,
  2003.

\bibitem{chao1}
A.~Chao, ``Nonparametric estimation of the number of classes in a population,''
  \emph{Scandinavian Journal of Statistics}, vol.~11, no.~4, pp. 265--270,
  1984.

\bibitem{chao2}
------, ``Estimating the population size for capture-recapture data with
  unequal catchability,'' \emph{Biometrics}, vol.~43, no.~4, pp. 783--791,
  1987.

\bibitem{sampleSurvey}
R.~K. Colwell, A.~Chao, N.~J. Gotelli, S.-Y. Lin, C.~X. Mao, R.~L. Chazdon, and
  J.~T. Longino, ``Models and estimators linking individual-based and
  sample-based rarefaction, extrapolation and comparison of assemblages,''
  \emph{Journal of Plant Ecology}, vol.~5, no.~1, p.~3, 2012.

\bibitem{stubborn}
X.~Yao, M.~Harman, and Y.~Jia, ``A study of equivalent and stubborn mutation
  operators using human analysis of equivalence,'' in \emph{Proceedings of the
  36th International Conference on Software Engineering}, ser. ICSE 2014, 2014,
  pp. 919--930.

\bibitem{csmith}
X.~Yang, Y.~Chen, E.~Eide, and J.~Regehr, ``Finding and understanding bugs in c
  compilers,'' in \emph{Proceedings of the 32Nd ACM SIGPLAN Conference on
  Programming Language Design and Implementation}, 2011, pp. 283--294.

\bibitem{randoop}
C.~Pacheco, S.~K. Lahiri, M.~D. Ernst, and T.~Ball, ``Feedback-directed random
  test generation,'' in \emph{Proceedings of the 29th International Conference
  on Software Engineering}, ser. ICSE '07, 2007, pp. 75--84.

\bibitem{afl}
Website, ``Afl: American fuzzy lop fuzzer,''
  \url{http://lcamtuf.coredump.cx/afl/technical\_details.txt}, 2017, accessed:
  2017-11-13.

\bibitem{aflfast}
M.~B\"{o}hme, V.-T. Pham, and A.~Roychoudhury, ``Coverage-based greybox fuzzing
  as markov chain,'' in \emph{Proceedings of the 2016 ACM SIGSAC Conference on
  Computer and Communications Security}, ser. CCS '16, 2016, pp. 1032--1043.

\bibitem{aflgo}
M.~B\"{o}hme, V.-T. Pham, M.-D. Nguyen, and A.~Roychoudhury, ``Directed greybox
  fuzzing,'' in \emph{Proceedings of the 24th ACM Conference on Computer and
  Communications Security}, ser. CCS, 2017, pp. 1--16.

\bibitem{aflsmart}
V.-T. Pham, M.~B\"{o}hme, A.~E. Santosa, A.~R. Caciulescu, and A.~Roychoudhury,
  ``Smart greybox fuzzing,'' \url{https://arxiv.org/abs/1811.09447}, 2018.

\bibitem{rare}
M.~Falk, J.~H{\"u}sler, and R.-D. Reiss, \emph{Laws of small numbers: extremes
  and rare events}.\hskip 1em plus 0.5em minus 0.4em\relax Springer Science \&
  Business Media, 2010.

\bibitem{evt}
E.~Castillo, \emph{Extreme value theory in engineering}.\hskip 1em plus 0.5em
  minus 0.4em\relax Elsevier, 2012.

\bibitem{rare1}
P.~Glasserman, P.~Heidelberger, P.~Shahabuddin, and T.~Zajic, ``Multilevel
  splitting for estimating rare event probabilities,'' \emph{Operations
  Research}, vol.~47, no.~4, pp. 585--600, 1999.

\bibitem{thinair}
Website, ``{AFL: Pulling Jpegs out of Thin Air, Michael Zalewski},''
  \url{https://lcamtuf.blogspot.com/2014/11/pulling-jpegs-out-of-thin-air.html},
  2017, accessed: 2017-11-13.

\bibitem{ace}
A.~Chao and S.-M. Lee, ``Estimating the number of classes via sample
  coverage,'' \emph{Journal of the American Statistical Association}, vol.~87,
  no. 417, pp. 210--217, 1992.

\bibitem{ice}
S.-M. Lee and A.~Chao, ``Estimating population size via sample coverage for
  closed capture-recapture models,'' \emph{Biometrics}, vol.~50, no.~1, pp.
  88--97, 1994.

\bibitem{goodtheory}
A.~Chao, C.-H. Chiu, R.~K. Colwell, L.~F.~S. Magnago, R.~L. Chazdon, and N.~J.
  Gotelli, ``Deciphering the enigma of undetected species, phylogenetic, and
  functional diversity based on good-turing theory,'' \emph{Ecology}, vol.~98,
  no.~11, pp. 2914--2929, 2017.

\bibitem{conroy}
M.~J. Conroy, J.~P. Runge, R.~J. Barker, M.~R. Schofield, and C.~J. Fonnesbeck,
  ``Efficient estimation of abundance for patchily distributed populations via
  two-phase, adaptive sampling,'' \emph{Ecology}, vol.~89, no.~12, pp.
  3362--3370, 2008.

\bibitem{thomson}
W.~Thompson, \emph{Sampling rare or elusive species: concepts, designs, and
  techniques for estimating population parameters}.\hskip 1em plus 0.5em minus
  0.4em\relax Island Press, 2013.

\bibitem{oracle}
E.~J. Weyuker, ``On testing non-testable programs,'' \emph{The Computer
  Journal}, vol.~25, no.~4, pp. 465--470, 1982.

\bibitem{asan}
K.~Serebryany, D.~Bruening, A.~Potapenko, and D.~Vyukov, ``Addresssanitizer: A
  fast address sanity checker,'' in \emph{Proceedings of the 2012 USENIX
  Conference on Annual Technical Conference}, 2012, pp. 28--28.

\bibitem{uncert}
S.~Elbaum and D.~S. Rosenblum, ``Known unknowns: Testing in the presence of
  uncertainty,'' in \emph{Proceedings of the 22Nd ACM SIGSOFT International
  Symposium on Foundations of Software Engineering}, ser. FSE 2014, 2014, pp.
  833--836.

\bibitem{coleman}
B.~D. Coleman, ``On random placement and species-area relations,''
  \emph{Mathematical Biosciences}, vol.~54, no.~3, pp. 191 -- 215, 1981.

\bibitem{cluster}
S.~K. Thompson, ``Adaptive cluster sampling,'' \emph{Journal of the American
  Statistical Association}, vol.~85, no. 412, pp. 1050--1059, 1990.

\bibitem{horvitz}
D.~G. Horvitz and D.~J. Thompson, ``A generalization of sampling without
  replacement from a finite universe,'' \emph{Journal of the American
  Statistical Association}, vol.~47, no. 260, pp. 663--685, 1952.

\end{thebibliography}

\end{document}